\documentclass[preprint]{aastex}

\usepackage{mathtext,bbm,amsmath,amsfonts,amssymb,indentfirst,syntonly,graphicx}
\usepackage[english]{babel}
\usepackage{slashbox}
\usepackage{calc}
\usepackage{tikz}
\usepackage{latexsym,graphicx,bbm}

\newcommand{\beq}{\begin{eqnarray}}
\newcommand{\eeq}{\end{eqnarray}}

\begin{document}

\title{Gamma-ray polarization of synchrotron-self-Compton process\\ from a highly relativistic jet}
\author{Zhe Chang\altaffilmark{1,2}, Hai-Nan Lin\altaffilmark{1,*}}
\affil{\altaffilmark{1}Institute of High Energy Physics\\Chinese Academy of Sciences, 100049 Beijing, China}
\affil{\altaffilmark{2}Theoretical Physics Center for Science Facilities\\Chinese Academy of Sciences, 100049 Beijing, China}
\altaffiltext{*}{linhn@ihep.ac.cn}

\begin{abstract}
The high polarization observed in the prompt phase of some gamma-ray bursts (GRBs) arouses extensive studies on the emission mechanism. In this paper, we investigate the polarization properties of the synchrotron-self-Compton (SSC) process from a highly relativistic jet. A magnetic-dominated, baryon-loaded jet ejected from the central engine travels with a large Lorentz factor. Shells with slightly different velocities collide with each other and produce shocks. The shocks accelerate electrons to power-law distribution, and at the same time, magnify the magnetic field. Electrons move in the magnetic field and produce synchrotron photons. The synchrotron photons suffer from the Compton scattering (CS) process and then are detected by an observer locating slightly off-axis. We derive analytically the formulae of photon polarization in the SSC process in two magnetic configurations: magnetic field in the shock plane and perpendicular to the shock plane. We show that photons induced by the SSC process can be highly polarized, with the maximum polarization $\Pi \sim 24\%$ in the energy band $[0.5,5]$ MeV. The polarization depends on the viewing angles, peaking in the plane perpendicular to the magnetic field. In the energy band $[0.05,0.5]$ MeV, in which most $\gamma$-ray polarimeters are active, the polarization is about twice of that in the Thomson limit, reaching to $\Pi\sim 20\%$. This implies that the Klein-Nishina effect, which is often neglected in literatures, should be carefully considered.
\end{abstract}

\section{Introduction}\label{sec:introduction}
Recent polarimetric observations in the prompt phase \citep{Coburn:2003,McGlynn:2007,Kalemci:2007,Gotz:2009,Yonetoku:2011,Yonetoku:2012}, as well as in the optical afterglow \citep{Bersier:2003,Steele:2009,Uehara:2012,Mundell:2013}, of some gamma-ray bursts (GRBs) show that the photons are highly linearly polarized. The circular polarization is often small enough to be ignored, especially in the afterglow phase \citep{Matsumiya:2003a}. For example, \citet{Wiersema:2012} reported the upper limit of circular polarization ($\Pi_{\rm cir}<0.15\%$) in the optical afterglow of GRB 091018. A little higher circular polarization ($\Pi_{\rm cir}=0.61\%\pm 0.13\%$) was found in the optical afterglow of GRB 121024A \citep{Wiersema:2014}. The first highly linear polarization was detected in the prompt phase of GRB 021206, which has a polarization of $80\%\pm 20\%$ \citep{Coburn:2003}. \citet{Kalemci:2007} reported the polarization as high as $98\%\pm 33\%$ in GRB 041219A, making it to be the most highly polarized GRB that has ever been observed, despite the low statistical significance. \citet{McGlynn:2007} re-checked the data of GRB 041219A, and found that the polarization is anti-correlated with photon energy. The analysis on the data of GRB 110301A and 110721A also shows certainly polarized, with polarization $70\%\pm 22\%(3.7\sigma)$ and $84_{-28}^{+16}\%(3.3\sigma)$, respectively \citep{Yonetoku:2012}. The temporal variabilities of polarization have also been detected in some GRBs, such as GRB 041219A \citep{Gotz:2009} and 100826A \citep{Yonetoku:2011}. Especially, the polarization angle of GRB 100828A shows a change of $\sim 90^{\circ}$ between two adjacent time intervals \citep{Yonetoku:2011}. In spite of large uncertainties exist, high polarization in the prompt phase of GRBs is still possible. The next generation gamma-ray polarimeter POLAR\footnote{http://polar.ihep.ac.cn/cms/.} on board the Chinese Space Laboratory Tian-Gong II is expected to provide more polarimetric data with unprecedented precision.

There are many theoretical interpretation for the origin of polarization. One of the most promising mechanisms to produce highly polarized photons is the synchrotron radiation. For isotropic electrons whose energies follow the power-law distribution (with index $p$), the maximum polarization of synchrotron photons is well-known to be $\Pi_{\rm syn}=(p+1)/(p+7/3)$, if the magnetic field is globally uniform \citep{Rybicki:1979}. The random magnetic field results much smaller polarization \citep{Gruzinov:1999a}. Another alternative mechanism is the Compton scattering (CS) process. The polarization of an unpolarized photon scattered by a static electron, in the Thomson approximation, is $\Pi_{\rm comp}=(1-\cos^2\theta_{\rm sc})/(1+\cos^2\theta_{\rm sc})$, where $\theta_{\rm sc}$ is the scattering angle of the photon \citep{Rybicki:1979}. Although photons can be completely polarized at the specific angle $\theta_{\rm sc}=90^{\circ}$, the probability is small since the cross section is minimum at this angle. \citet{Lazzati:2004} investigated the polarization properties of an isotropic photon field scattered by an electron jet (the so-called Compton drag model), and showed that the polarization, in some special cases, can be as large as that in the point-source limit. However, they only discussed in the Thomson limit, and the seed photons were assumed to be unpolarized. In the energy bands in which most $\gamma$-ray polarimeters are active (e.g., $[50, 500]$ keV), the Thomson approximation may be invalid, and the study of polarization in the Klein-Nishina region is quite necessary. On the other hand, the seed photons may originate from synchrotron radiation, thus are initially polarized. \citet{Toma:2009} used the Monte Carlo simulation to derive the polarization of photons in three emission models, and found that high polarization is possible. The ICMART model proposed by \citet{Zhang:2011} can well reproduce the variabilities of the polarization degree observed in some GRBs, but can not naturally interpret the change of polarization angle in GRB 100826A. Except for the above intrinsic mechanisms, the polarization can also arise from geometric effects \citep{Granot:2003,Waxman:2003}. In the random magnetic field case, high polarization is still possible if the jet opening angle is small enough and the line-of-sight is close to the edge of the jet.

The emission region may be optically thick, thus the synchrotron photons may suffer from photon-electron scattering. Therefore, the study of photon polarization in the synchrotron-self-Compton (SSC) process is necessary. Actually, the polarization properties of the SSC process for an isotropic electron distribution have already been calculated analytically in the Thomson approximation \citep{Celotti:1994,Poutanen:1994}. On the other hand, the Monte Carlo simulations have also been used by some authors to calculate the polarization of the SSC process \citep{McNamara:2009,Krawczynski:2012}. In some special configurations, polarization of $\sim 50\%$ can be easily realized. However, an analytical calculation of polarization of the SSC process in the Klein-Nishina region is still lacking. In recent papers \citep{Chang:2013,Chang:2014a,Chang:2014b}, we have presented analytically, a general formulism to calculate the polarization properties of an initially polarized photon scattered by electrons with any energy distribution (e.g., power-law electrons). The photon polarization depends on the photon energy and viewing angle. It was showed that the photon-electron scattering process can produce a wide range of polarization, ranging from completely unpolarized, to completely polarized. The predicted polarization-energy relation is coincident with the observation on GRB 041219A. If certain conditions are satisfied, the change of polarization angle observed in the prompt phase of GRB 100826A can be naturally interpreted. The magnetic-dominated jet model predicts that, for long GRBs, MeV photons are emitted during the acceleration phase \citep{Chang:2012}. If this is true, the change of polarization angle is a natural result of the expansion of the jet \citep{Chang:2014c}.

Based on the previous works, we investigate, in this paper, the $\gamma$-ray polarization in the SSC process from a highly relativistic jet. The SSC process is a natural prediction of the magnetic-dominated jet model \citep{Meszaros2011,Veres:2012sb}. According to this model, a highly relativistic and magnetized jet, which contains shells with slightly different velocities, ejects from the central engine. Different Shells collide with each other and produce shocks. The shocks accelerate electrons to be power-law distribution, and at the same time, magnify the magnetic field. Electrons moving in the uniform magnetic field radiate synchrotron photons, which followed by photon-electron collision before escaping from the jet. Starting from the differential cross section of photon-electron scattering, we analytically derive the polarization of a photon after being scattered by any electron. Then we integrate over the spectra of electrons and photons, thus the polarization of the SSC process is obtained. We will show that, for isotropic and power-law electrons, photons induced by the SSC process can have high net polarization. The Klein-Nishina effect contributes significantly to the polarization of the prompt emission of GRBs.

The rest of the paper is arranged as follows. Section \ref{sec:compton} is devoted to a short review on the photon polarization in the CS process. In section \ref{sec:syn-compton}, we calculate the polarization properties of the SSC process from a highly relativistic jet in two magnetic configurations. Finally, discussions and conclusions are given in section \ref{sec:conclusion}.

\section{Compton scattering process}\label{sec:compton}
In this section, we will give a short review on the polarization properties of the CS process. This section includes two subsections. In subsection \ref{sec:single-scatter}, we consider the process of a photon scattered by an electron moving with any velocity, while subsection \ref{sec:isotropic} deals with the isotropic electron case. The formulae presented in this section are valid not only in the Thomson region, but also in the  Klein-Nishina region.

\subsection{A photon scattered by an electron}\label{sec:single-scatter}
Suppose a photon with energy $\varepsilon_0$ collides with an electron traveling with any velocity. In the laboratory frame, the electron initially  goes along the $\hat{\mathbf l}_0$ direction. After being scattered, the photon is scattered to the $\hat{\mathbf n}$ direction. We set a Cartesian coordinate system in the laboratory frame, such that its origin is fixed to the collision point, the $z$-axis is along the direction of incident photon, the $y$-axis is in the scattering plane (the plane in which the incident and scatted photons are contained), and $\hat{\mathbf x}=\hat{\mathbf y}\times\hat{\mathbf z}$. In this coordinate system, the moving direction of the scattered photon can be written as
\begin{equation}
  \hat{\mathbf n}=\sin\theta_{\rm sc}~\hat{\mathbf y} + \cos\theta_{\rm sc}~\hat{\mathbf z},
\end{equation}
where $\theta_{\rm sc}$ is the scattering angle, i.e., the angle between the directions of incident and scattered photons. The direction of the incident electron can be written as
\begin{equation}
  \hat{\mathbf l}_0=\sin\theta_2\cos\varphi_2~\hat{\mathbf x} + \sin\theta_2\sin\varphi_2~\hat{\mathbf y} + \cos\theta_2~\hat{\mathbf z},
\end{equation}
where $\theta_2$ and $\varphi_2$ are the polar and azimuthal angles of the incident electron, respectively. The conservation of the four-momentum gives a constraint on the energy of the scattered photon, that is \citep{Akhiezer:1965},
\begin{equation}\label{eq:energy}
 \varepsilon_1=\frac{\varepsilon_0(1-\beta_e \cos\theta_2)}{\frac{\varepsilon_0}{\gamma_e m_e c^2}(1-\cos\theta_{\rm sc})+(1-\beta_e\cos\theta_1)},
\end{equation}
where $\gamma_e$ is the Lorentz factor of the incident electron, $\beta_e=\sqrt{1-1/\gamma_e^2}$\, is the velocity of the electron, and $\theta_1$ is the angle between $\hat{\mathbf n}$ and $\hat{\mathbf l}_0$.

The polarization of the scattered photon can be deduced starting from the differential cross section of photon-electron scattering. The polarization-dependent cross section of photon-electron scattering, is given in the laboratory frame as \citep{Berest:1982,Chang:2014a,Chang:2014b}
\begin{equation}\label{eq:cross-section}
  d \sigma = \frac{1}{4} r_e^2 d \Omega \left(\frac{\varepsilon_1}{\varepsilon_0}\right)^2  \bigg[ F_0 +F_3(\xi_3+\xi'_3) + F_{11} \xi_1 \xi_1' +F_{22} \xi_2\xi'_2+F_{33} \xi_3\xi'_3\bigg],
\end{equation}
where $r_e=e^2/m_ec^2$ is the classical electron radius, $d\Omega=\sin\theta d\theta d\varphi$, and the quantities $F_a$ $(a=0,3,11,22,33)$ are functions of the initial and final states of the photon and electron. The expression of $F_a$ can be found at \citet{Chang:2014a,Chang:2014b}.

The Stokes parameters, $\xi_i$ and $\xi'_i$ ($i=1,2,3$) in Eq.(\ref{eq:cross-section}), stand for the polarization properties of the incident and scattered photons, respectively. The linear polarization of a photon can be completely described by $\xi_3$ and $\xi_1$. The remaining parameter, $\xi_2$, is a representation for the circular polarization, positive for right-handed and negative for left-handed. As is mentioned in the introduction, the circular polarization of GRB photons is small, so we will ignore it in this paper. The polarization of the incident photon, $\Pi_0$, can be conveniently written in terms of the Stokes parameters as
\begin{equation}\label{eq:pi0}
 \Pi_0=\sqrt{(\xi_1)^2+(\xi_2)^2+(\xi_3)^2}.
\end{equation}
On the contrary, given a photon with linear polarization $\Pi_0$, we can write its Stokes parameters as
\begin{equation}
  \xi_1=\Pi_0\sin 2\chi_0,~~\xi_2=0,~~\xi_3=\Pi_0\cos 2\chi_0,
\end{equation}
where the polarization angle $\chi_0\in[-\pi/2,\pi/2]$ is the angle between the polarization vector and the $x$-axis in the $xy$ plane.

From the quantum electrodynamics, we can derive the Stokes parameters of the secondary photon. They are given as \citep{Berest:1982}
\begin{equation}\label{eq:stokes}
 \xi^{\rm f}_1=\frac{ \xi_1 F_{11}}{F_0+\xi_3 F_3}, \quad
 \xi^{\rm f}_2=\frac{ \xi_2 F_{22}}{F_0+\xi_3 F_3}, \quad
 \xi^{\rm f}_3=\frac{F_3+ \xi_3 F_{33}}{F_0+\xi_3 F_3}.
\end{equation}
The secondary photon is still circularly unpolarized if the incident photon is circularly unpolarized. In particular, if the incident photon is unpolarized (neither linearly nor circularly), we have $\xi_1^{\rm f}=\xi_2^{\rm f}=0$, and $\xi_3^{\rm f}=F_3/F_0$. In the electron rest case, $\xi_3^{\rm f}$ is positive definitely, which means that the polarization vector of the secondary photon is always perpendicular to the scattering plane. This is a well-known property of the CS process. Similar to the incident photon, the polarization of the secondary photon, $\Pi$, can be derived from Eq.(\ref{eq:pi0}) by replacing $\xi_i$ with $\xi^{\rm f}_i$.

\subsection{A photon scattered by isotropic electrons}\label{sec:isotropic}
In this subsection, we consider the polarization of a photon after being scattered by isotropic electrons. We assume that the energy of the electrons follows the power-law distribution, i.e., ${\mathcal N}_e(\gamma_e)d\gamma_e \propto \gamma_e^{-p}d\gamma_e$. Once the single CS process is studied clearly, the polarization of a photon scattered by isotropic electrons can be easily obtained by averaging over the energy and orientational angle of the incident electrons.

Similar to the single scattering case, the Stokes parameters of a photon after being scattered by isotropic electrons, $\langle\xi^{\rm f}_i\rangle$, can be derived from Eq.(\ref{eq:stokes}) by replacing $F_a$ with the averaged components $\langle F_a\rangle$, which are defined to be \citep{Chang:2014a,Chang:2014b}
\begin{equation}\label{eq:Fa_averaged}
 \langle F_a (\varepsilon_0,\theta_{\rm sc})\rangle\equiv\frac{1}{C}\int_{\gamma_1}^{\gamma_2}{\cal N}_e(\gamma_e) d\gamma_e\int_0^{\pi}\sin\theta_2 d\theta_2\int_0^{2\pi}d\varphi_2\left(\frac{\varepsilon_1}{\varepsilon_0}\right)^2 F_a.
\end{equation}
Here, $C\equiv\int_{\gamma_1}^{\gamma_2}{\cal N}_e(\gamma_e)d\gamma_e\int_0^{\pi}\sin\theta_2d\theta_2\int_0^{2\pi}d\varphi_2$ is the normalization factor. It is a nuisance factor, since the stokes parameters of the scattered photon are the ratios of the linear combination of $\langle F_a\rangle$ (see Eq.(\ref{eq:stokes})), thus the normalization factor is completely canceled out. In Eq.(\ref{eq:Fa_averaged}), we have assumed that the Lorentz factors of the incident electrons are in the range $\gamma_e\in[\gamma_1,\gamma_2]$. Therefore, the polarization of the photon after being scattered by isotropic electrons, is given as
\begin{equation}\label{eq:pi-scattered}
 \langle\Pi(\varepsilon_0,\theta_{\rm sc})\rangle=\sqrt{\langle\xi^{\rm f}_1\rangle^2+\langle\xi^{\rm f}_2\rangle^2+\langle\xi^{\rm f}_3\rangle^2}.
\end{equation}

In the Thomson limit, i.e., $\varepsilon_0\ll m_ec^2$, the above formulae are much simpler. In the leading order approximation, Eq.(\ref{eq:energy}) is reduced to
\begin{equation}
\varepsilon_1=\frac{\varepsilon_0(1-\beta_e \cos\theta_2)}{1-\beta_e\cos\theta_1}.
\end{equation}
The polarization of the photon after scattering is simplified to \citep{Chang:2014b}
\begin{equation}\label{eq:pi-thomson}
  \langle\Pi(\theta_{\rm sc})\rangle=\Pi_0\frac{\langle F_{11} \rangle}{\langle F_0 \rangle},
\end{equation}
where
\begin{equation}
  F_0=\frac{\varepsilon_1}{\varepsilon_0}+\frac{\varepsilon_0}{\varepsilon_1}-\sin^2\theta_{\rm sc},
\end{equation}
and
\begin{equation}\label{eq:Sigma}
 F_{11}=\frac{2}{\gamma_e^2(1-\beta_e\cos\theta_2)^2}\left(1-\frac{\beta_e\sin\theta_2\sin\varphi_2}{1-\beta_e\cos\theta_2} \tan\frac{\theta_{\rm sc}}{2}\right).
\end{equation}
We can see that the polarization of the scattered photon is independent of photon energy and initial polarization angle. Differing from the single scattering case, the scattered photon is polarized only if the incident photon is polarized. This is because the isotropic distribution of the electrons cancels out the polarization. Nevertheless, if the incident photon is initially polarized (e.g., synchrotron photon), the net polarization of the scattered photon cannot be completely canceled out.

\section{SSC process from a highly relativistic jet}\label{sec:syn-compton}
In this section, we calculate the polarization of photons in the SSC process from a highly relativistic jet. Consider a highly relativistic, magnetic-dominated, and baryon-loaded jet ejected from the central engine travels with a large Lorentz factor. Shells with slightly different velocities collide with each other and produce shocks. The shocks accelerate electrons to power-law distribution, and at the same time, magnify the magnetic field. Electrons move in the uniform magnetic field and produce synchrotron photons. The synchrotron photons collide with the seed electrons and then escape from the jet, and are detected by the observer at a certain direction. For simplicity, we assume that both the synchrotron cooling and Compton cooling are slow enough such that the electrons always keep the power-law distribution. We consider two different cases which have been extensively studied in literatures: (1) magnetic field in the shock plane, and (2) magnetic field perpendicular to the shock plane. These two globally ordered magnetic configurations can be advected by the jet from the central engine \citep{Spruit:2001,Fendt:2004}.

\subsection{Magnetic field in the shock plane}\label{sec:parallel}
\begin{figure}[htbp]
\centering
  \includegraphics[width=9 cm]{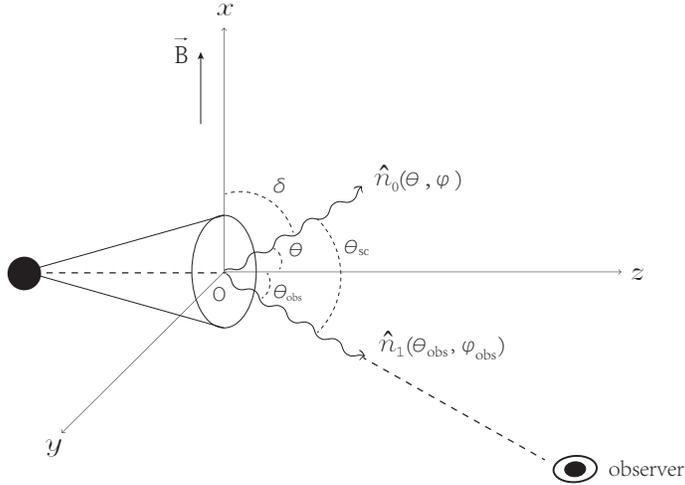}
  \caption{\small{A schematic representation of the SSC process in the case that the magnetic field is contained in the shock plane. We choose a Cartesian coordinate system such that the $z$-axis is in the jet direction, the $x$-axis is parallel to the magnetic field, and the $xyz$ axes form the right-handed set.}}\label{fig:Geometry2}
\end{figure}

Fig.\ref{fig:Geometry2} shows the case that the magnetic field is contained in the shock plane.
Consider a photon with energy $\varepsilon_0$ travels along any direction $\hat{\mathbf n}_0$ and collides with isotropic electrons. After scattering, the photon goes towards the $\hat{\mathbf n}_1$ direction and then is detected by an observer. We set a Cartesian coordinate system in the jet comoving frame, such that the $z$-axis is in the jet direction, the $x$-axis is parallel to the magnetic field, and $\hat{\mathbf y}=\hat{\mathbf z}\times\hat{\mathbf x}$. The directions of the incident and scattered photons can be written as
\begin{eqnarray}
    \hat{\mathbf n}_0&=&\sin\theta\cos\varphi~\hat{\mathbf x}+\sin\theta\sin\varphi~\hat{\mathbf y}+\cos\theta~\hat{\mathbf z},\\
    \hat{\mathbf n}_1&=&\sin\theta_{\rm obs}\cos\varphi_{\rm obs}~\hat{\mathbf x}+\sin\theta_{\rm obs}\sin\varphi_{\rm obs}~\hat{\mathbf y}+\cos\theta_{\rm obs}~\hat{\mathbf z},
\end{eqnarray}
where $\theta$ and $\varphi$ ($\theta_{\rm obs}$ and $\varphi_{\rm obs}$) are the polar and azimuthal angles of the incident (scattered) photon, respectively. The scattering angle is given as
\begin{eqnarray}\nonumber
  \cos\theta_{\rm sc}=\hat{\mathbf n}_0\cdot\hat{\mathbf n}_1&=&\sin\theta\cos\varphi\sin\theta_{\rm obs}\cos\varphi_{\rm obs}\\
  &&+\sin\theta\sin\varphi\sin\theta_{\rm obs}\sin\varphi_{\rm obs}+\cos\theta\cos\theta_{\rm obs}.
\end{eqnarray}

If the electrons are isotropic and their energies follow the power-law distribution $({\mathcal N}_e(\gamma_e)d\gamma_e\propto \gamma_e^{-p}d\gamma_e)$, the spectrum of synchrotron photons is also power-law, but not isotropic. Most photons are emitted in the plane perpendicular to the magnetic field \citep{Rybicki:1979},
\begin{equation}
  N_{\gamma}(\varepsilon_0,\theta,\varphi)\propto\varepsilon_0^{-\frac{p-1}{2}}(\sin\delta)^{\frac{p+1}{2}},
\end{equation}
where $\delta$ is the pitch angle of the synchrotron photon with respect to the direction of the magnetic field (the $x$-axis in our case), that is,
\begin{equation}
  \cos\delta=\hat{\mathbf x}\cdot\hat{\mathbf n}_0=\sin\theta\cos\varphi.
\end{equation}
The polarization of the photons induced by the SSC process can be easily obtained by averaging over the photon spectrum, i.e.,
\begin{equation}
  \langle\langle\Pi(\theta_{\rm obs},\varphi_{\rm obs})\rangle\rangle=\frac{\int\langle\Pi(\varepsilon_0,\theta_{\rm sc})\rangle N_{\gamma} (\varepsilon_0,\theta,\varphi) \sin\theta d\theta d\varphi d\varepsilon_0}{\int N_{\gamma} (\varepsilon_0,\theta,\varphi) \sin\theta d\theta d\varphi d\varepsilon_0},
\end{equation}
where $\langle\Pi(\varepsilon_0,\theta_{\rm sc})\rangle$ is given by Eq.(\ref{eq:pi-scattered}), or in the Thomson limit, by Eq.(\ref{eq:pi-thomson}).

Up to now, we are working in the jet comoving frame. The jet moves towards the observer highly relativistically. In order to transform the above formulae to the observer frame, the following relation is useful,
\begin{equation}\label{eq:theta}
  \cos\theta_{\rm obs}=\frac{\cos\bar{\theta}_{\rm obs}-\beta_{\rm jet}}{1-\beta_{\rm jet}\cos\bar{\theta}_{\rm obs}},~~
  \varphi_{\rm obs}=\bar{\varphi}_{\rm obs},
\end{equation}
where $\beta_{\rm jet}=(1-1/\Gamma^2)^{1/2}$ is the velocity of the jet with respect to the observer in unit of light speed, and $\Gamma$ is the Lorentz factor of the jet. Hereafter, the quantities in the observer frame are denoted with a bar. Since the polarization is Lorentz invariant \citep{Cocke:1972}, we have
\begin{equation}\label{eq:Pi-obs}
  \langle\langle\bar{\Pi}(\bar{\theta}_{\rm obs},\bar{\varphi}_{\rm obs})\rangle\rangle=\langle\langle\Pi(\theta_{\rm obs},\varphi_{\rm obs})\rangle\rangle.
\end{equation}

\begin{figure}[htbp]
\centering
  \includegraphics[width=12 cm]{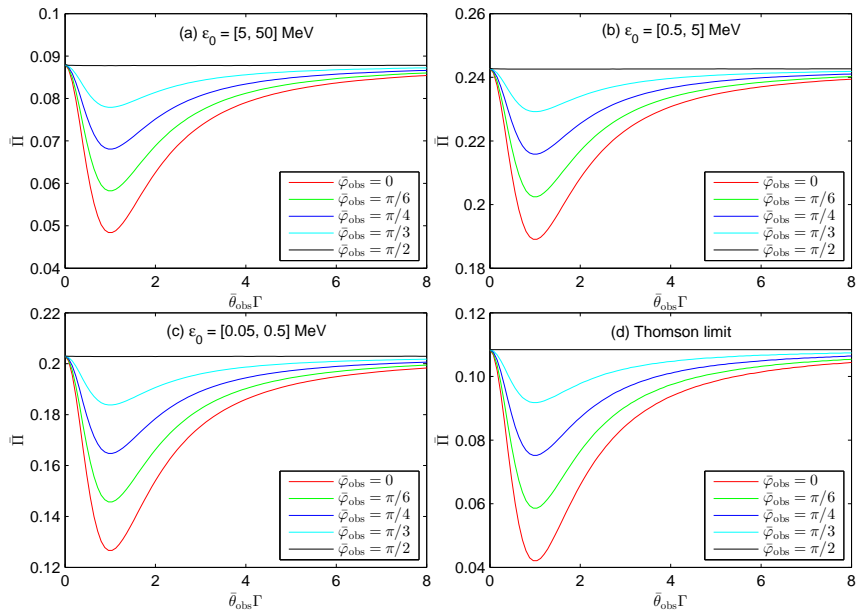}
  \caption{\small{The polarization of photons as a function of viewing angles in the case that the magnetic field is contained in the shock plane. The polarization in four energy bands are showed in different panels.}}\label{fig:PI1_4panels.eps}
\end{figure}

Fig.\ref{fig:PI1_4panels.eps} shows the evolution of the polarization with the viewing angles $\bar{\theta}_{\rm obs}$ in four different energy bands: $\varepsilon_0=[5,50]$, $[0.5,5]$, $[0.05,0.5]$ MeV, and in the Thomson limit.
Different curves stand for different azimuthal angles $\bar{\varphi}_{\rm obs}$. In the plot, we have taken the index of electron spectrum to be $p=3$. Therefore, the polarization of the synchrotron photons is about $\Pi_0\approx 75\%$. The jet Lorentz factor is set to be $\Gamma\approx 200$ \citep{Chang:2012}. We have assumed that the Lorentz factors of incident electrons are in the range $\gamma_e\in [1,10]$. The contribution to polarization from electrons with larger Lorentz factors is negligible \citep{Chang:2014a,Chang:2014b}. The polarization angle $\chi_0$ has been averaged. As is showed in Fig.\ref{fig:PI1_4panels.eps}, the polarization of photons in the energy band $\varepsilon_0=[0.5,5]$ MeV is larger than that in any other bands. This is because the polarization as a function of photon energy peaks at about $\varepsilon_0\sim 1$ MeV \citep{Chang:2014a}. In the Thomson limit, the maximum polarization is about 10\%. In the energy band $\varepsilon_0=[0.05,0.5]$ MeV, in which most $\gamma$-ray polarimeters are active, the maximum polarization is about twice of that in the Thomson limit. This implies that the Klein-Nishina effect should be considered in the theoretical calculations of the $\gamma$-ray polarization. For each energy band, the polarization reaches its maximum at $\bar{\varphi}_{\rm obs}=\pi/2$, corresponding to the plane perpendicular to the magnetic field. In this plane, the polarization is independent of $\bar{\theta}_{\rm obs}$. Another noticeable feature is that, for a fixed azimuthal angle $\bar{\varphi}_{\rm obs}$, the polarization has a minimum at $\bar{\theta}_{\rm obs}\Gamma\approx 1$. In the jet frame, this angle corresponds to $\theta_{\rm obs}\approx \pi/2$, i.e., the shock plane ($xy$ plane). Generally speaking, the observer whose line-of-sight is perpendicular (parallel) to the magnetic field sees the highest (lowest) polarization.

\subsection{Magnetic field perpendicular to the shock plane}\label{sec:perpendicular}
\begin{figure}[htbp]
\centering
  \includegraphics[width=9 cm]{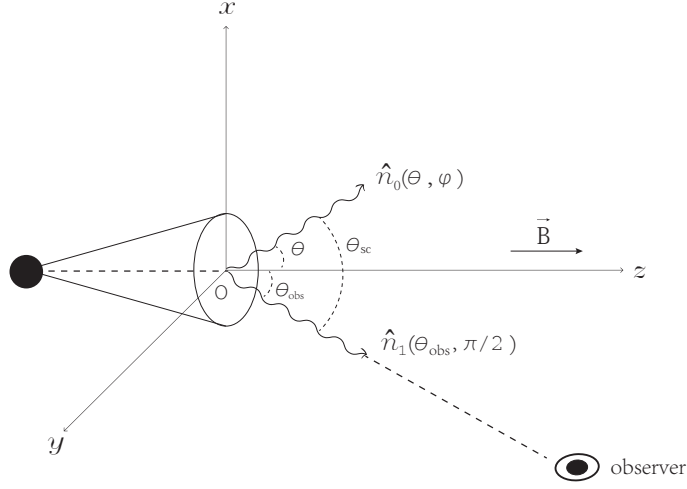}
  \caption{\small{A schematic representation of the SSC process in the case that the magnetic field is perpendicular to the shock plane. We choose a Cartesian coordinate system such that the $z$-axis is in the jet direction, the $y$-axis is in the plane which contains the directions of jet and line-of-sight, and the $xyz$ axes form the right-handed set.}}\label{fig:Geometry3}
\end{figure}

This subsection is similar to the last subsection, except that the magnetic field is perpendicular to the shock plane (see Fig.\ref{fig:Geometry3}).
We set a Cartesian coordinate system in the jet comoving frame, such that the $z$-axis is in the jet direction, the $y$-axis is in the plane which contains the directions of jet and line-of-sight, and $\hat{\mathbf x}=\hat{\mathbf y}\times\hat{\mathbf z}$. In such a coordinate system, the directions of incident and scattered photons can be written as
\begin{eqnarray}
    \hat{\mathbf n}_0&=&\sin\theta\cos\varphi~\hat{\mathbf x}+\sin\theta\sin\varphi~\hat{\mathbf y}+\cos\theta~\hat{\mathbf z},\\
    \hat{\mathbf n}_1&=&\sin\theta_{\rm obs}~\hat{\mathbf y}+\cos\theta_{\rm obs}~\hat{\mathbf z},
\end{eqnarray}
respectively, and the scattering angle is given as
\begin{equation}
  \cos\theta_{\rm sc}=\hat{\mathbf n}_0\cdot\hat{\mathbf n}_1=\sin\theta\sin\varphi\sin\theta_{\rm obs}+\cos\theta\cos\theta_{\rm obs}.
\end{equation}

For isotropic and power-law $({\mathcal N}_e(\gamma_e)d\gamma_e\propto \gamma_e^{-p}d\gamma_e)$ electrons, the spectrum of synchrotron photons is \citep{Rybicki:1979}
\begin{equation}
  N_{\gamma}(\varepsilon_0,\theta)\propto\varepsilon_0^{-\frac{p-1}{2}}(\sin\theta)^{\frac{p+1}{2}}.
\end{equation}
After integrating over the photon spectrum, we obtain the polarization of the SSC process as a function of viewing angle $\theta_{\rm obs}$, that is,
\begin{equation}
  \langle\langle\Pi(\theta_{\rm obs})\rangle\rangle=\frac{\int\langle\Pi(\varepsilon_0,\theta_{\rm sc})\rangle N_{\gamma} (\varepsilon_0,\theta) \sin\theta d\theta d\varphi d\varepsilon_0}{\int N_{\gamma} (\varepsilon_0,\theta) \sin\theta d\theta d\varphi d\varepsilon_0}.
\end{equation}
Note that in this case, due to the axis-symmetry, the polarization is independent of the azimuthal angle $\varphi_{\rm obs}$. With the help of Eqs.(\ref{eq:theta}) and (\ref{eq:Pi-obs}), we can transform the polarization from the jet frame to the observer frame.

\begin{figure}[htbp]
\centering
  \includegraphics[width=12 cm]{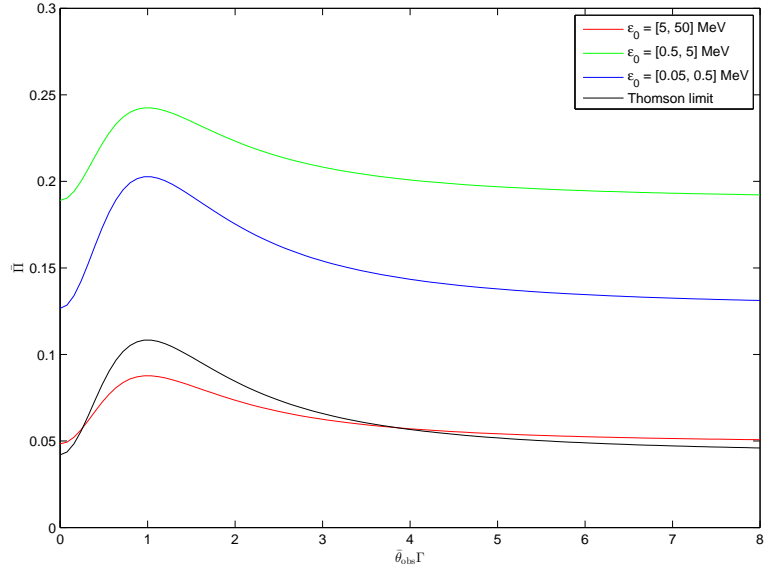}
  \caption{\small{The polarization of photons as a function of viewing angles in the case that the magnetic field is perpendicular to the shock plane. Different lines stands for different energy bands.  Note that in this case, the polarization is independent of $\bar{\varphi}_{\rm obs}$.}}\label{fig:PI2_4panels.eps}
\end{figure}

In Fig.\ref{fig:PI2_4panels.eps}, we plot the polarization of the SSC photons as a function of viewing angle $\bar{\theta}_{\rm obs}$ in four energy bands: $\varepsilon_0=[5,50]$, $[0.5,5]$, $[0.05,0.5]$ MeV, and in the Thomson limit.
In the plot, the parameters are chosen to be the same to the last subsection, i.e., $p=3$, $\Gamma=200$ and $\gamma_e\in [1,10]$. As was done in the last subsection, we have also averaged over the polarization angle $\chi_0$. Similar to the last case, photons in the energy band $\varepsilon_0=[0.5,5]$ MeV have the highest polarization, which can be as high as 24\%. Photons with energy much higher than 50 MeV are almost unpolarized (which is not plotted in this figure). The polarization in the energy band $\varepsilon_0=[0.05,0.5]$ MeV is approximately twice of that in the Thomson limit. In each energy band, the polarization has a peak at the viewing angle $\bar{\theta}_{\rm obs}\Gamma\approx 1$, corresponding to $\theta_{\rm obs}\approx\pi/2$ in the jet frame, i.e., the plane perpendicular to the magnetic field ($xy$ plane).

\section{Discussions and conclusions}\label{sec:conclusion}
In this paper, we have presented an analytical calculation of $\gamma$-ray polarization induced by the SSC process from a highly relativistic jet in the Klein-Nishina region. We investigated the scenario that isotropic electrons radiate synchrotron photons in the globally uniform magnetic field. The synchrotron photons are scattered by the seed electrons and then escape from the jet. After integrating over the electron distribution and photon spectrum, the polarization of the SSC process was obtained. We calculated two cases which have been extensively discussed in literatures: (1) magnetic field in the shock plane, and (2) magnetic field perpendicular to the shock plane. These two globally uniform magnetic configurations, although cannot be produced by the shock, can be advected by the jet from the central engine. Instead of firstly working in the electron-rest frame, then transforming to the jet frame, as was done by most authors, in this paper we directly work in the jet frame. This has the advantage of avoiding the complicate Lorentz transformation between these two frames. The general formulae presented in this paper are valid in the Klein-Nishina region, as well as in the Thomson region. The formulism is useful in calculating the polarization of the SSC process in astrophysical processes, such as GRBs and AGNs.

We numerically calculated the polarization of the SSC process from isotropic electrons with power-law distribution. We found that, photons induced by the SSC process can be highly polarized, despite that the seed electrons are isotropic. In both magnetic configurations, photons in the energy band $\varepsilon_0=[0.5,5]$ MeV have the highest polarization, reaching to about 24\%. This is due to the fact that the polarization peaks at $\varepsilon_0\sim 1$ MeV. Photons with energy much higher than 50 MeV are almost unpolarized. The polarization of photons in the energy band $\varepsilon_0=[0.05,0.5]$ MeV (most polarimetric observations in the prompt phase of GRBs are performed in this energy band) is about 20\%, twice of that in the Thomson limit. This implies that the Klein-Nishina effect, which is often neglected in literatures, should be carefully considered. In both magnetic configurations, the observer whose line-of-sight is perpendicular to the magnetic field sees the highest polarization. On the contrary, if we see along the direction of the magnetic field, we see the lowest polarization. Magnetic field plays an important role in the polarization effect. If isotropic photons are scattered by isotropic electrons, the polarization certainly vanishes due to the symmetry. The existence of magnetic field breaks the symmetry of photon distribution, thus the polarization arises.

The temporal variabilities of polarization observed in GRB 041219A and GRB 100826A may be partially due to the evolution of the bulk Lorentz factor $\Gamma$. Especially, if the jet accelerates smoothly (e.g., the magnetic-dominated jet model predicts that $\Gamma\propto r^{1/3}$), the polarization angle can be changed $90^{\circ}$ suddenly at a critical value of $\Gamma$ \citep{Chang:2014c}. Therefore, the observation of polarization may help us to learn the evolution process of the outflow. Moreover, recent studies on $\gamma$-ray polarization in the photospheric emission models and some comptonized emission models show the anti-correlation between the polarization degree and the luminosity \citep{Lundman:2014}. Interestingly, in the case that the magnetic field is contained in the shock plane, we find that the polarization degree increases as the viewing angle increases, if photons are emitted at a large viewing angle ($\bar{\theta}_{\rm obs}\Gamma\gtrsim 1$) The larger viewing angle leads to the lower luminosity. Thus, bright GRBs show low polarization, while dimmer ones show high polarization. However, in the case that the magnetic field is perpendicular to the shock plane, the situation is completely opposite. The future observation of polarization-luminosity relation provides a way to distinguish the magnetic configurations in the emission region.

We stress that the real astrophysical process is much more complicate than the scenarios investigated in this paper. Firstly, we regarded the emission region as a point source. For a typical GRB of redshift $z\sim 1$, it locates at a distance about Gpc away from us. This distance is much larger than the size of emission region, thus the point-source limit is a good approximation. Secondly, we only considered the slow cooling case, i.e., we assumed that the radiation is slow enough such that the electrons and photons keep their original spectra. Otherwise, after radiating synchrotron photons, the spectrum of electrons is changed, and this further affects the photon spectrum. This correlated process heavily complicates the calculation, or even makes it impossible. Thirdly, for simplicity, we only considered the single-scattering process. In fact, multi-scattering processes often take place. After one scattering process, the polarization and spectrum of photons are all changed. Such a complicate situation can only be studied using the Monte Carlo simulations. Finally, the $e^+e^-$ pair production and annihilation above the threshold may also affect the polarization significantly. All of these issues remain to be the interesting future work.

\begin{acknowledgments}
We are grateful to X. Li, P. Wang, S. Wang and D. Zhao for useful discussion. This work has been funded by the National Natural Science Fund of China under Grant No. 11375203.
\end{acknowledgments}

\end{document}